# Selective Addressing of Coupled Qubits via Complex Frequency Zero Targeting

Deepanshu Trivedi[1], Laraib Niaz[1], and Alex Krasnok[1,2]

[1]Department of Electrical Engineering, Florida International University, 33174, Miami, USA

[2]Knight Foundation School of Computing and Information Sciences, Florida International University, 33174, Miami, USA

To whom correspondence should be addressed: akrasnok@fiu.edu

## Abstract

Achieving precise, individual control over qubits within scalable quantum processors is critically hampered by parasitic couplings and spectral crowding, leading to detrimental crosstalk. While optimal absorption strategies based on time-reversal symmetry have shown promise for single emitters, their applicability is limited in realistic multi-qubit systems where realistic losses break time-reversal symmetry. This work introduces a robust approach using complex frequency (CF) pulses specifically tailored to the complex reflection zeros of the complete, coupled, and explicitly lossy qubit-waveguide system. This method circumvents the limitations of idealized time-reversal arguments by directly engaging with the dissipative system's true response characteristics. We first develop a theoretical framework for a system of three coupled two-level emitters, employing Heisenberg equations to derive the system's response and design appropriate CF pulses that inherently account for the system's dissipative nature. The efficacy and practicality of this approach are then validated through comprehensive transient simulations for a realistic model of three Josephson junction-based transmon qubits, explicitly including intrinsic qubit losses. Our results demonstrate that CF pulses can selectively excite a target qubit with significantly suppressed crosstalk to neighboring qubits, markedly outperforming conventional Gaussian pulses of comparable energy.

## 1. Introduction

The pursuit of fault-tolerant quantum computation has spurred intensive research into developing scalable quantum hardware and sophisticated control techniques capable of manipulating quantum states with high precision[1–6]. Superconducting qubits, due to their design flexibility, compatibility with microwave control, and potential for integration using established semiconductor fabrication techniques, have emerged as a leading platform for building quantum processors[7–11]. A fundamental requirement for any quantum computing architecture is the ability to initialize, manipulate, and read out the state of individual qubits with high fidelity. As these systems scale to larger numbers of qubits, achieving such precise control becomes increasingly challenging.

A prevalent architectural paradigm in superconducting quantum processors involves coupling multiple qubits to a shared bus, such as a microwave resonator or a transmission line waveguide. This shared infrastructure serves as a common conduit for delivering control signals and extracting readout information, facilitating connectivity and reducing the sheer number of required control lines[4,12–16]. While advantageous for scalability in terms of wiring density, this shared-bus architecture introduces significant complexities. As qubit density increases, their resonance frequencies can become closely spaced, leading to spectral crowding. This proximity, compounded by both direct parasitic couplings (capacitive or inductive) between neighboring qubits and indirect couplings mediated by the shared bus, results in unwanted inter-qubit interactions and control crosstalk[8,17–22]. Crosstalk manifests as the unintentional driving of non-target qubits when a control pulse is applied to a specific target qubit, which significantly degrades the fidelity of quantum operations. Effectively addressing these crosstalk challenges is paramount for the realization of scalable and high-performance quantum computers. For instance, typical pulse lengths of tens of nanoseconds for single-qubit gates result in drive pulse bandwidths in the tens of megahertz range, which can easily overlap with closely spaced qubit frequencies. This issue is exacerbated by the inherent ~1-2% fabrication imprecision in qubit frequencies, which necessitates post-fabrication tuning or advanced control techniques to avoid detrimental frequency collisions.

Conventional approaches to qubit control often employ microwave pulses with well-defined carrier frequencies, typically shaped by Gaussian or other smooth envelopes, to drive transitions in target qubits[23–28]. The selectivity of such pulses fundamentally relies on their spectral bandwidth being sufficiently narrow to predominantly address only the intended qubit. However, in systems with closely spaced qubit frequencies or strong inter-qubit couplings, the spectral tails of these pulses can significantly overlap with the transitions of non-target qubits. This leads to off-resonant excitation and, consequently, substantial crosstalk, hindering high-fidelity operations. While various advanced pulse shaping techniques have been developed to mitigate these effects, the challenge of achieving high selectivity in densely coupled systems persists as an active area of research. Prominent among these are methods like DRAG (Derivative Removal by Adiabatic Gate), which aims to reduce leakage to higher excited states and can be designed to minimize off-resonant driving by carefully shaping the in-phase and quadrature components of the control pulse[28–31]. Optimal control theory (OCT), often implemented via algorithms like GRAPE (Gradient Ascent Pulse Engineering), offers a powerful numerical framework to design complex pulse shapes that maximize gate fidelity while respecting system constraints, including the suppression of crosstalk[21,32–36]. These methods, however, can result in complex pulse shapes that may be challenging to interpret or implement experimentally due to bandwidth limitations of arbitrary waveform generators. More recently, techniques like Selective Excitation Pulses (SEP) have been proposed, which shape a drive pulse to create null points in its frequency spectrum at the frequencies of non-target qubits[28]. This approach, demonstrated experimentally for three fixed-frequency transmon qubits sharing a control line, has shown single-qubit gate fidelities comparable to conventional Gaussian pulses while effectively suppressing unwanted excitations. The SEP method focuses on engineering the frequency domain amplitude of the pulse to achieve selectivity. Another class of approaches involves active crosstalk cancellation schemes or the use of dynamical decoupling (DD) sequences applied to spectator qubits to effectively isolate the target qubit during gate operations by averaging out unwanted interactions. For instance, the XY4 DD sequence has been experimentally shown to suppress ZZ crosstalk and improve both quantum memory and gate performance on IBM quantum processors[37].

Similarly, methods based on understanding and leveraging the system's input-output characteristics, such as achieving *coherent perfect absorption* or *impedance matching* to specific modes via time-reversal symmetry, are being explored[38–41]. Approaches relying on time-reversed waveforms for optimal absorption, while highly effective for single emitters in near-ideal conditions, face limitations in realistic scenarios. Specifically, the inherent dissipative losses present in any real quantum system break the strict time-reversal symmetry of the Hamiltonian. This means the poles and zeros of the system's scattering matrix (e.g., reflection coefficient) are no longer necessarily complex conjugate pairs, and a simple time-reversed emission profile of an isolated qubit or a frequency derived from naive conjugation relationships valid only for lossless systems may not be optimal for absorption when that qubit is part of a lossy, coupled network. Moreover, extending such single-emitter arguments directly to multi-qubit systems presents a considerable challenge.

In this work, we explore and systematically develop an alternative yet complementary approach based on complex frequency (CF) pulses[42] that is directly applicable to realistic, dissipative multi-qubit systems. The merit of our approach lies in recognizing that the system's response, including losses, is encapsulated by its complex poles and zeros. The core idea is to meticulously shape the excitation pulse by precisely tuning both its carrier frequency (real part) and its temporal envelope, specifically its decay rate (imaginary frequency component). This allows the pulse to be critically "impedance matched" not to a simplified, lossless model, but to a specific complex pole (a zero of reflection) of the entire coupled, and potentially lossy, qubit-waveguide system[42–46]. When an input signal is engineered to this precise complex frequency, it can achieve near-perfect reflectionless regime, even in the presence of intrinsic dissipation that breaks time-reversal symmetry. We explicitly demonstrate that this requires targeting the true reflection zeros of the lossy system, as deviations based on, for example, conjugating the poles (a strategy that might be inferred from lossless system properties) lead to suboptimal performance, including increased reflection and crosstalk. More significantly, this energy is then channeled into exciting a particular collective eigenmode of the system, which, through careful system design and pole selection, can be made to predominantly involve the target qubit. This approach promises not only highly efficient energy transfer to the desired qubit but also the simultaneous minimization of reflections and, crucially, the active suppression of excitation pathways to other qubits through precisely engineered destructive interference effects. This moves beyond simple spectral avoidance or idealized time-reversal arguments to a more fundamental matching with the system's actual, complex response characteristics, offering a novel and robust pathway to enhanced quantum control in realistic multi-qubit architectures.

We systematically investigate the application of CF pulses for achieving high-fidelity selective excitation in a multi-qubit system. Our study begins with the development of a theoretical model based on temporal input-output formalism for a system comprising three distinct two-level quantum emitters (qubits) coupled to a single-mode resonator, which is, in turn, coupled to an external waveguide[47–49]. By solving the linearized Heisenberg equations of motion, we derive the reflection coefficient of the system and identify its complex poles. These poles then inform the design of the CF pulses. Subsequently, we translate this concept to a more experimentally relevant scenario by simulating a system of three Josephson junction (JJ) based transmon qubits coupled to a common microwave transmission line. These simulations are performed using Keysight's Advanced Design System (ADS), a powerful commercial microwave circuit simulator, incorporating a realistic nonlinear model for JJs and characterized intrinsic qubit losses. Our findings consistently show that CF pulses, tailored to the specific complex reflection zeros of the system, can achieve a remarkable degree of selective excitation even in these dissipative conditions. This is quantitatively substantiated by Target Selectivity and Crosstalk Suppression Ratio metrics, which reveal marked improvements for CF pulses over conventional methods and resilience in lossy environments when true system zeros are targeted. Both the idealized theoretical model and the detailed ADS simulations for JJ qubits demonstrate a significant reduction in crosstalk to non-targeted qubits compared to conventional Gaussian pulses of similar energy. This enhanced selectivity and efficiency, particularly the method's success in explicitly lossy circuits where naive extensions of time-reversal symmetry or lossless system properties fail, highlights the potential of CF pulse shaping as a valuable tool for precise quantum control in complex, integrated quantum circuits.

## 2. Theoretical Model and System Dynamics

This section outlines the theoretical framework developed to illustrate the core principles of selective qubit excitation using complex frequency pulses. To provide an intuitively clear demonstration, we employ a Bloch equation approach for analyzing the dynamics of the hybrid quantum system and deriving an analytical expression for its reflection coefficient. This treatment, describing the evolution of the expectation values of the qubit and resonator observables, allows for a transparent understanding of the underlying physics. While this simplified model, focusing on two-level approximations for the qubits, does not capture all complexities of realistic superconducting qubits, such as higher energy levels, intricate loss mechanisms, or specific nonlinearities of Josephson junctions, it serves to effectively demonstrate the fundamental concept of pole-zero excitation for achieving selectivity. The insights gained from this illustrative Bloch-like framework are subsequently shown to be applicable and are validated by the ADS simulations of Josephson junction qubits presented later in this work.

The system under consideration for this illustrative model comprises three spatially separated qubits. These qubits are each coupled to a single-mode resonator, which in turn is connected to a one-dimensional waveguide. The waveguide serves a dual function as both the conduit for incoming excitation pulses and the output channel for the scattered field. The analysis leverages the principles of waveguide QED in combination with the input-output formalism. The configuration, illustrated in Fig. 1(a), involves three qubits coupled to a resonator mode. For clarity and to isolate the effect of frequency detuning, all qubits are assumed to have identical coupling strengths, $g$, with $g$ chosen as $\omega_c/100$. The resonator and the first qubit (Q1) are frequency-aligned at $\omega_c = \omega_{a1} = 2\pi \times 5.77$ GHz. The second (Q2) and third (Q3) qubits are detuned by 1% and 2% relative to $\omega_c$, respectively. This corresponds to detunings of approximately 57.7 MHz for Q2 ($\omega_{a2} = 1.01\omega_c$) and 115.4 MHz for Q3 ($\omega_{a3} = 1.02\omega_c$). These frequency separations are representative of state-of-the-art multi-qubit processors where careful frequency allocation is employed to manage addressability while accommodating fabrication tolerances.

In this illustrative model, we consider the resonator to be an ideal, lossless system. This assumption implies that there are no internal energy dissipation mechanisms, which can be formally stated by setting the rate of internal losses, $1/\tau_i$, to zero, where $\tau_i$ represents the characteristic time for such internal decay processes. The resonator's interaction with the external one-dimensional waveguide is characterized by a radiative decay rate, denoted as $1/\tau_r$. Here, $\tau_r$ is the characteristic time associated with the energy leaking from the resonator into the waveguide. The sharpness of the resonator's response due to this coupling is quantified by its quality factor, $Q$, which for this lossless resonator is determined by its coupling to the waveguide and is fixed at $Q = 31$. The system is excited by an input field, represented as the time-dependent amplitude $b_{in}(t)$, which propagates from the waveguide towards the resonator. This incident field couples to the resonator mode at a specific coupling rate, $k_r$, which is related to the radiative decay time by the expression $k_r = \sqrt{2/\tau_r}$.

System dynamics are formulated in the Heisenberg picture. The resonator state is captured by the expectation value of its annihilation operator, $a(t)$, while the $j$-th qubit is described by the lowering operator $\sigma_-^{(j)}(t)$ and population inversion operator $\sigma_z^{(j)}(t)$. The time evolution of these quantities follows a coupled set of nonlinear differential equations. The equation for the resonator field is:

$$\frac{d}{dt}\langle a(t)\rangle = \left(i\omega_c - \frac{1}{\tau_r}\right)\langle a(t)\rangle + k_r b_{in}(t) - i\sum_{j=1}^{3} g_j \left\langle \sigma_-^{(j)}(t)\right\rangle \quad (1)$$

For each qubit $j = 1,2,3$, the equations are:

$$\frac{d}{dt}\langle \sigma_-^{(j)}(t)\rangle = \left(i\omega_{aj} - \frac{\Gamma_s^{(j)}}{2}\right)\langle \sigma_-^{(j)}(t)\rangle + ig_j\langle a(t)\rangle\left\langle \sigma_z^{(j)}(t)\right\rangle \quad (2)$$

$$\frac{d}{dt}\left\langle \sigma_z^{(j)}(t)\right\rangle = -\Gamma_s^{(j)}\left(\left\langle \sigma_z^{(j)}(t)\right\rangle + 1\right) + 2ig_j\left(\langle a^\dagger(t)\rangle\langle \sigma_-^{(j)}(t)\rangle - \langle a(t)\rangle\left\langle \sigma_+^{(j)}(t)\right\rangle\right) \quad (3)$$

Here, $\Gamma_s^{(j)}$ denotes the spontaneous emission rate of qubit $j$ into non-resonator modes. Nonlinearities arise from terms like $\langle a(t)\rangle\langle\sigma_z^{(j)}(t)\rangle$, capturing the dependence of the system's response on qubit state.

The system dynamics, as formulated in the Heisenberg picture by Eqs. (1-3), can be rigorously derived from a full quantum mechanical treatment of the coupled qubit-resonator-waveguide system (see *Supplementary Materials S1* for details). The derivation begins with the total system Hamiltonian, $\hat{H}_{\text{total}}$, which encompasses the individual Hamiltonians for the qubits (treated as two-level systems with transition frequencies $\omega_{aj}$), the resonator (a harmonic oscillator mode with frequency $\omega_c$), and the waveguide (a continuum of bosonic modes), along with their interactions. Key interaction terms include a Tavis-Cummings type Hamiltonian describing the coupling between the qubits and the resonator mode (with strength $g_j$, typically derived under a rotating wave approximation, RWA), and terms describing the resonator-waveguide interaction as well as qubit and resonator coupling to other environmental modes leading to dissipation. The equations of motion for the system operators ($\hat{a}$, $\hat{\sigma}_-^{(j)}$, $\hat{\sigma}_z^{(j)}$) are then obtained using the Heisenberg equation, $d\hat{O}/dt = (1/i\hbar)[\hat{O}, \hat{H}_{\text{total}}]$. The crucial step of incorporating the waveguide's influence involves standard input-output theory, which, under the Born-Markov approximation[10,50–52], allows for the elimination of the waveguide bath operators. This procedure results in the resonator decay term ($1/\tau_r$) into the waveguide and introduces the input field operator $\hat{b}_{in}(t)$ that drives the resonator mode $\hat{a}$ at a rate $k_r = \sqrt{2/\tau_r}$. Similarly, the coupling of individual qubits to their respective local electromagnetic environments accounts for the spontaneous emission terms characterized by rates $\Gamma_s^{(j)}$. The Eqs. (1-3), are for the expectation values of these system operators (e.g., $\langle a(t)\rangle \equiv \langle \hat{a}(t)\rangle$). This step involves taking the expectation value of the operator Heisenberg-Langevin equations. The nonlinear terms, such as $\langle a(t)\rangle\langle\sigma_z^{(j)}(t)\rangle$, appear directly from the expectation value of operator products without invoking a mean-field factorization at this stage, thus retaining a more accurate description of the qubit-resonator interaction dynamics.

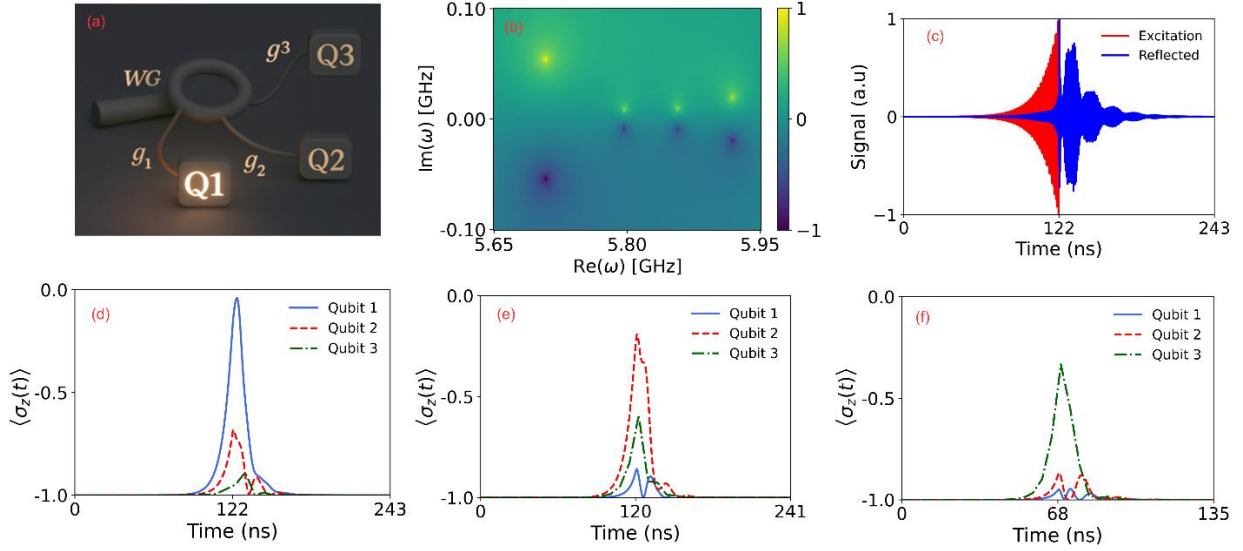

**Figure 1.** Theoretical demonstration of selective qubit excitation using CF pulses. (a) Schematic of the hybrid quantum system: three qubits (Q1, Q2, Q3) coupled with strength $g$ to a single-mode resonator, which is coupled to a one-dimensional waveguide (WG). (b) Magnitude of the reflection coefficient $r(\omega)$ calculated from Eq. (5) in the complex frequency plane ($\omega = \omega_r - i\omega_i$). Dark spots indicate the reflection zeros of the system, which define the target parameters for CF pulse design. (c) Example time-domain profile of an input CF pulse (red) tuned to a system pole, and the corresponding greatly diminished reflected signal (blue), indicating efficient energy absorption as predicted by $r(\omega) \approx 0$. (d)-(f) Selective qubit excitation dynamics showing the population inversion $\langle\sigma_z^{(j)}(t)\rangle$ for each qubit. Minimal crosstalk is observed for non-target qubits, validating the selectivity of the CF pulse approach within this theoretical model.

To derive an analytical solution and determine the characteristic complex frequencies, the equations are linearized under the weak excitation approximation, assuming the input field is sufficiently weak that all qubits remain near their ground states. This yields: $\langle\sigma_z^{(j)}(t)\rangle \approx -1$ for $j = 1,2,3$. This simplification enables frequency-domain analysis by assuming harmonic solutions with a complex frequency $\omega = \omega_r - i\omega_i$: $b_{in}(t) = b_0 e^{-i\omega t}$, $\langle a(t)\rangle = A e^{-i\omega t}$, $\langle\sigma_-^{(j)}(t)\rangle = \Sigma_-^{(j)} e^{-i\omega t}$. Substitution into equations (1) and (2) yields a linear algebraic system for the amplitudes $A$ and $\Sigma_-^{(j)}$: $\left(-i\omega + i\omega_c - \frac{1}{\tau_r}\right)A + i\sum_{j=1}^{3} g_j \Sigma_-^{(j)} = k_r b_0$, $\left(-i\omega + i\omega_{aj} - \frac{\Gamma_s^{(j)}}{2}\right)\Sigma_-^{(j)} - ig_j A = 0$. These can be compactly represented as $M \cdot v = s$, where $v = [A, \Sigma_-^{(1)}, \Sigma_-^{(2)}, \Sigma_-^{(3)}]^T$ and $s = [k_r b_0, 0, 0, 0]^T$. The system matrix $M$ is:

$$M = \begin{pmatrix} i\omega_c - i\omega - \frac{1}{\tau_r} & i\mathbf{g}^T \\ -i\mathbf{g} & \mathbf{D} \end{pmatrix}, \quad (4)$$

Where $\mathbf{g} = (g_1, g_2, g_3)^T$ is the coupling vector, $\mathbf{D}$ is a diagonal matrix: $\mathbf{D} = \text{diag}\left(i\omega_{a1} - i\omega - \frac{\Gamma_s^{(1)}}{2}, i\omega_{a2} - i\omega - \frac{\Gamma_s^{(2)}}{2}, i\omega_{a3} - i\omega - \frac{\Gamma_s^{(3)}}{2}\right)$. The reflected output is given by $b_{out}(t) = -b_{in}(t) + k_r \langle a(t) \rangle$, yielding the reflection coefficient, $r(\omega) = -1 + k_r \frac{A}{b_0}$. Solving for $A$ (e.g., using Cramer's rule: $A = \det(M_A)/\det(M)$), the full expression for $r(\omega)$ becomes:

$$r(\omega) = -1 - \frac{2}{\tau_r} \frac{\prod_{j=1}^{3}\left(i\omega - i\omega_{aj} + \frac{\Gamma_s^{(j)}}{2}\right)}{\det(M)} \quad (5)$$

The complex frequencies $\omega$ satisfying $r(\omega) = 0$ represent the system's transfer function zeros. At these frequencies, incoming waves are completely absorbed. Tailored excitation pulses centered at these frequencies can achieve efficient, selective energy transfer into the corresponding eigenmodes, which can be engineered to target a specific qubit preferentially.

Fig. 1(a) schematics of the coupled system: three qubits (Q1, Q2, Q3) identically coupled to a single-mode resonator, which is itself coupled to an input-output waveguide (WG). The complex frequency response of this system, as captured by the derived reflection coefficient $r(\omega)$, is depicted in Fig. 1(b). This plot shows the magnitude of $r(\omega)$ in the complex frequency plane ($\omega = \omega_r - i\omega_i$). The dark regions correspond to frequencies where $|r(\omega)| \approx 0$; these are the complex reflection zeros of the system. According to our model, an input pulse whose carrier frequency and temporal envelope are tailored to match one of these specific complex zeros will be efficiently absorbed by the system, ideally with no reflection. Panel (c) illustrates this principle in the time domain: an input pulse (red) designed as a complex frequency (CF) pulse targeting one such pole results in a significantly diminished reflected signal (blue), confirming efficient energy transfer as predicted by the vanishing reflection coefficient.

The goal of this tailored energy transfer is selective qubit excitation. Panels (d)-(f) of Fig. 1 present the simulated dynamics of the individual qubit populations, $\langle \sigma_z^{(j)}(t) \rangle$, when CF pulses are applied. These dynamics are solutions to the full nonlinear Heisenberg equations of motion (Eqs. 1-3), where the input field $b_{in}(t)$ is shaped as a CF pulse corresponding to one of the system's poles identified in panel (b). When the CF pulse is designed to target the eigenmode predominantly associated with Qubit 1 (panel d), Qubit 1 (solid blue line) exhibits significant excitation from its ground state towards an excited state, while Qubit 2 (dashed red line) and Qubit 3 (dotted green line) remain largely unexcited. Similarly, panel (e) shows the selective excitation of Qubit 2 when its corresponding eigenmode is targeted, and panel (f) demonstrates the selective excitation of Qubit 3. In all instances, the crosstalk to non-target qubits is minimal. These theoretical results, derived from the system's scattering properties and its fundamental equations of motion, strongly support the hypothesis that CF pulses, designed based on the system's complex zeros, can achieve high-fidelity, selective control of individual qubits in a coupled environment. For simulation results of conventional Gaussian pulse excitation within the theoretical model (Fig. S1), provided for comparison, see *Supplementary Materials (S2).*

## 3. JJ-Qubit System Implementation

To connect the idealized theoretical model with a more experimentally realistic scenario, we implemented and simulated a system of three coupled transmon qubits, each based on a JJ, using Keysight's Advanced Design System (ADS). ADS was used to model the detailed electromagnetic behavior of coupled superconducting transmon qubit systems, incorporating realistic nonlinear JJ models and enabling transient and frequency-domain analysis of their response to complex control pulses and inter-element coupling[53–59]. The simulated configuration, schematically shown in Figure 2(a), consists of three transmon qubits. Each transmon is modeled as a lumped-element circuit incorporating a JJ shunted by a parallel capacitor ($C_j$). The required nonlinearity for qubit behavior is provided by JJ, which functions as a nonlinear inductor. Its inductance $L_j$ varies with the junction current $I_j$ according to $L_j = L_{j0} / \sqrt{1 - (I_j / I_c)^2}$, where $L_{j0} = \phi_0 / (2\pi I_c)$ is the zero-bias inductance, $\phi_0 = h/(2e)$ is the magnetic flux quantum, and $I_c$ is the junction's critical current[4,8]. The transmon's large shunt capacitance ensures a high ratio between Josephson energy ($E_J = I_c \phi_0 / (2\pi)$) and charging energy ($E_C = e^2 / (2C_\Sigma)$, where $C_\Sigma$ is the total shunt capacitance), placing the system deep in the transmon regime ($E_J / E_C \gg 1$). This suppresses charge noise sensitivity and improves coherence times. The parameters used for ADS simulations were: JJ inductance (zero bias): $L_{j1} = L_{j2} = L_{j3} = 2$ nH; Junction capacitances: $C_{j1} = C_{j2} = C_{j3} = 0.2$ pF; Critical current for all qubits: $I_c = 0.1647$ µA; Resulting energy ratio: $E_J / E_C \approx 850$.

All three qubits are coupled to a common microwave transmission line (control bus). The coupling is capacitive, with the coupling capacitors intentionally detuned to produce distinguishable resonance frequencies when each qubit interacts with the bus. The coupling capacitors follow $C_{cj} = C_{ci}(1+\delta)^k$, with $\delta = 0.1$ and $k$ being an integer. Specifically: $C_{c1} = 10$ fF (Qubit 1), $C_{c2} = 11$ fF (Qubit 2, i.e., $C_{c1}(1+0.1)$), $C_{c3} = 12.1$ fF (Qubit 3, i.e., $C_{c1}(1+0.1)^2$). The control line connects to Port via a coupling capacitor $C_\kappa = 10$ pF. The eigenmode structure of the system, including anti-crossings that signal qubit-qubit coupling, is illustrated in Figure 2(b), with corresponding complex-plane reflection zeros shown in Figure 2(c).

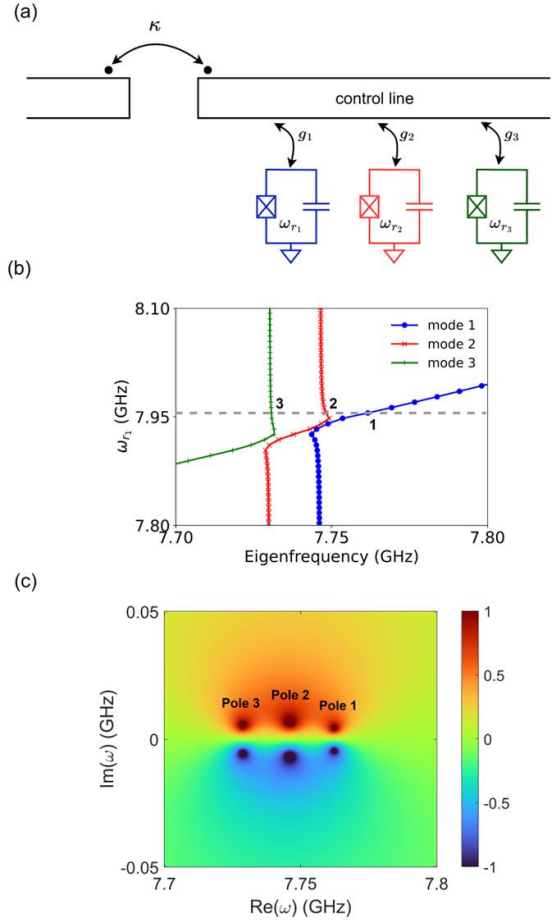

**Figure 2.** (a) Schematic of the simulated circuit consisting of three JJ-based transmon qubits, each with a nominal resonant frequency $\omega_{rj}$ and shunting capacitance $C_j$. The qubits are capacitively coupled ($C_{cj}$) to a common microwave control line, which is interfaced with an external port via a coupling capacitance $C_\kappa$. (b) Eigenfrequencies of the coupled three-qubit system simulated using the QuCAT Python library, plotted as a function of the bare frequency of Qubit 1 ($\omega_{r1}$). The three distinct curves represent the system's eigenmodes, exhibiting anti-crossings indicative of inter-qubit coupling. The dashed horizontal line marks the operational point chosen for the simulations, where the modes can be identified with predominant contributions from Qubit 1, Qubit 2, and Qubit 3 (labeled '1', '2', '3'). (c) Magnitude of the reflection coefficient, $|r(\omega)|$, plotted in the complex frequency plane ($\omega = \text{Re}(\omega) + i\text{Im}(\omega)$) for the system at the selected operational point. The dark blue regions (labeled Pole 1, Pole 2, Pole 3) represent the reflection zeros of the coupled system. These zeros provide the target complex frequencies for designing the CF pulses used for selective qubit excitation in the subsequent ADS transient simulations.

We conducted transient simulations in ADS to study the time-domain behavior of the coupled qubit system under microwave pulse excitation. The simulations utilized standard ADS library components, including a behavioral JJ model. The JJ was modeled using ADS's "JJ2" two-terminal component, based on the Resistively and Capacitively Shunted Junction (RCSJ) model, which allows for accurate simulations of superconducting microwave circuits. The supercurrent $I_j$ through the junction obeys the Josephson relation $I_j = I_c \sin(2\pi\phi/\phi_0)$, where $\phi$ is the phase difference across the junction. The Josephson Junction is a superconducting device which exhibits zero resistance when the absolute value of the current flowing through the device is less than critical current. To ensure fidelity with control schemes derived from linearized dynamics and to keep the qubits in the weakly anharmonic regime, excitation power was kept low. This maintains $I_j \ll I_c$, typically below 5% of $I_c$, thereby approximating linearized behavior.

The JJ2 model includes parameters such as the shunt resistance ($R_{shunt}$) and McCumber parameter ($\beta_C$). For primary simulations, these parameters were left at their default values (e.g., effectively infinite $R_{shunt}$, $\beta_C = 0$), indicating no explicit shunt resistor across the junction in the model. While real devices may incorporate shunt resistors to suppress hysteresis, transmons typically operate with junctions voltage-biased via the microwave drive and capacitive network, not current-biased.

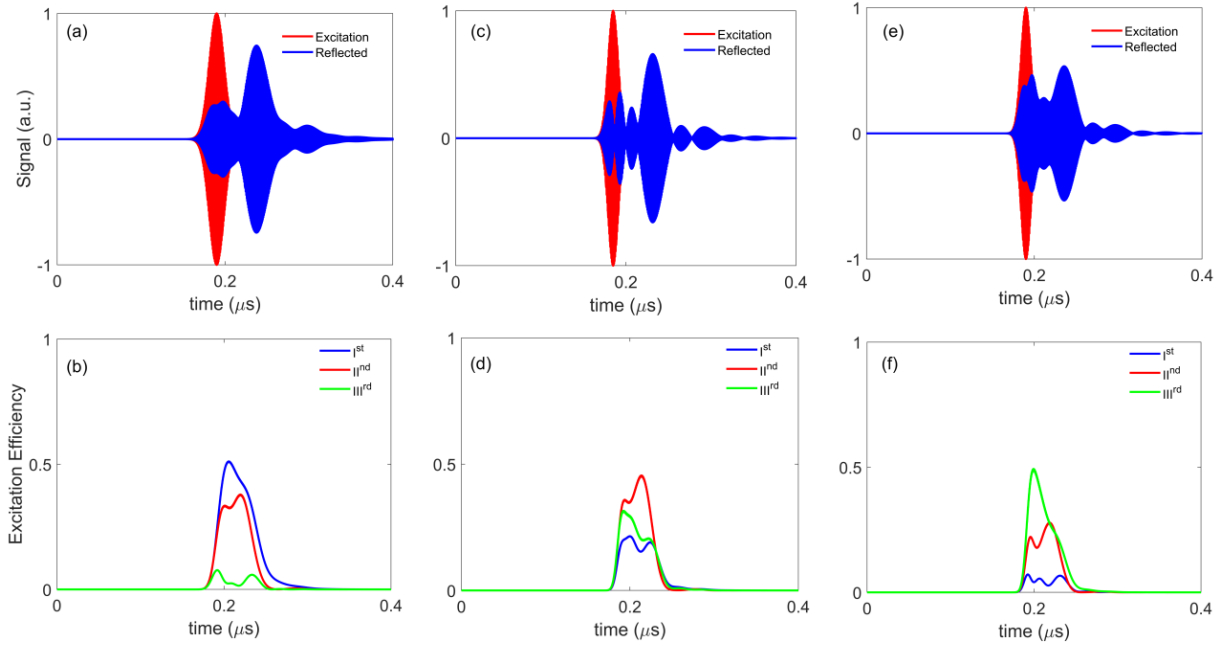

**Figure 3.** Qubit excitation dynamics using conventional Gaussian pulses. The panels show the system response when targeting Qubit 1 (a, b), Qubit 2 (c, d), and Qubit 3 (e, f) with Gaussian pulses tuned to their respective real eigenfrequencies. (a), (c), (e): Time-domain profiles of the input Gaussian pulse (red) and the corresponding reflected signal (blue) from the control line. (b), (d), (f): Corresponding Excitation Efficiency for Qubit 1 (blue solid line), Qubit 2 (red solid line), and Qubit 3 (green solid line). While the target qubit is excited, significant unwanted excitation (crosstalk) is observed in the non-target qubits in all cases.

An essential aspect of simulating nonlinear JJ circuits is selecting an appropriate time step. The initial time step was chosen to resolve the highest frequency components relevant to the qubit transitions and the drive. To further optimize accuracy and numerical stability, we employed the build-in Random Optimization controller in Keysight ADS. In addition to fine tuning component values, we also optimized the simulation time step (set to $t_0 = 0.001$ ns ). The Random Optimizer minimizes the average weighted deviation from the desired responses using a least-squares error function or mean squared error (MSE), making it well-suited for optimizing parameters in systems with complex, time-domain behaviors. The optimization goal was to minimize reflected power at the end of the excitation pulse This is critical for nonlinear reactive components like JJs, where numerical accuracy and convergence are highly sensitive to the integration step size. A well-chosen time step ensures fidelity without excessively long runtimes or numerical instability. This level of time step optimization is generally unnecessary in linear circuit simulations, which are less sensitive to such variations.

To quantitatively assess and compare the performance of different control strategies, we define the excitation efficiency, $\eta(t)$. This metric represents the instantaneous ratio of the energy stored in the targeted resonator mode, proportional to $|\langle a(t)\rangle|^2$, to the total energy injected by the excitation signal, $s_{\text{ex}}(t)$, integrated up to time $t$:

$$\eta(t) = \frac{|\langle a(t)\rangle|^2}{\int_0^t |s_{\text{ex}}(t')|^2\, dt'} \cdot 100\% \quad (6)$$

This definition allows for a time-resolved measure of how effectively the input pulse energy is being converted into excitation of the desired mode within the system.

First, we examine the system's response to a standard control methodology using Gaussian-enveloped microwave pulses. For each selective excitation attempt, the carrier frequency of the Gaussian pulse was tuned to match the real part of the eigenfrequency of the target qubit's mode, as identified in the analysis shown in Figure 2(b). The simulation results for targeting Qubit 1, Qubit 2, and Qubit 3 are compiled in Figure 3. A key observation from the time-domain plots in Figures 3(a), 3(c), and 3(e) is the significant signal reflected from the input port. The amplitude of the reflected pulse is a substantial fraction of the input pulse amplitude, which indicates inefficient power transfer from the control line into the desired qubit mode. This impedance mismatch arises because a simple spectrally-shaped pulse does not account for the complex admittance of the coupled multi-qubit system. The Gaussian pulse is defined with a pulse width $\sigma_i$, but the effective pulse width (measured between points where the amplitude drops to $1/\sqrt{2}$ of the peak) is approximately $1.62\sigma_i$. Additionally, the energy in both Gaussian and complex frequency pulses is comparable.

The detrimental consequence of this approach is most evident in the excitation efficiency plots, shown in Figures 3(b), 3(d), and 3(f). When targeting Qubit 1 (panel b), although it reaches the highest excitation level, there is significant crosstalk to Qubit 2, which reaches an excitation efficiency of approximately 35% relative to the target. Similarly, targeting Qubit 2 (panel d) results in substantial crosstalk to both Qubit 1 and Qubit 3.

The attempt to excite Qubit 3 (panel f) also shows considerable unwanted excitation of its neighbors. This demonstrates that for a spectrally crowded system with strong inter-qubit coupling, conventional frequency-selective Gaussian pulses fail to provide adequate addressability, leading to errors that would severely limit the fidelity of any quantum algorithm.

Next, we evaluated the performance of the CF pulse technique. Each CF pulse was designed with a carrier frequency (real part) and a temporal decay rate (imaginary part) that precisely match one of the complex zeros of the system identified in Figure 2(c). The results of these simulations are presented in Figure 4. The improvement is immediate and striking. The time-domain plots in Figures 4(a), 4(c), and 4(e) show a dramatically suppressed reflected signal. The near-complete absorption of the input pulse signifies that the CF pulse is effectively impedance-matched to the system, allowing for highly efficient energy transfer into the targeted eigenmode.

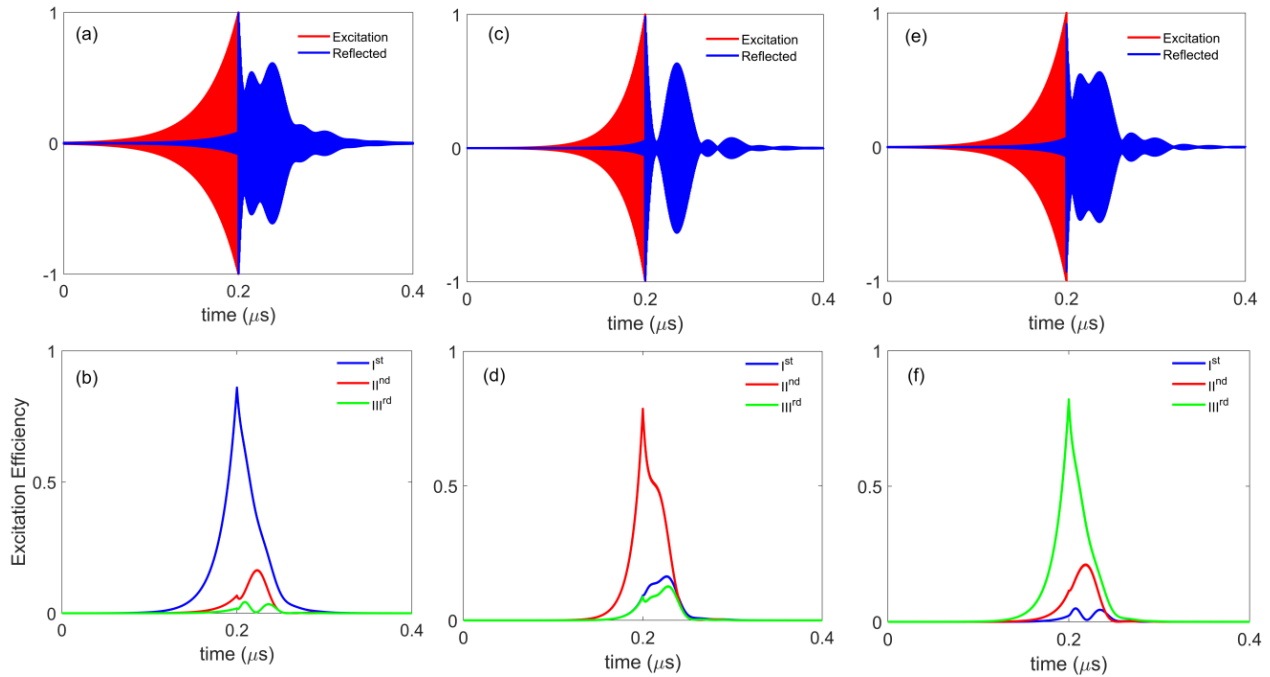

**Figure 4**. High-fidelity selective qubit excitation using CF pulses. The panels show the system response when targeting Qubit 1 (a, b), Qubit 2 (c, d), and Qubit 3 (e, f) with tailored CF pulses. (a), (c), (e): Time-domain profiles of the input CF pulse (red) and the corresponding reflected signal (blue) from the control line. The significantly reduced amplitude of the reflected signal indicates efficient energy absorption by the system. (b), (d), (f): Corresponding Excitation

Efficiency for Qubit 1 (blue solid line), Qubit 2 (red solid line), and Qubit 3 (green solid line). Each plot demonstrates high selectivity, with the target qubit being strongly excited while excitation of the non-target qubits is minimal, showcasing suppressed crosstalk.

This efficient energy transfer is coupled with exceptional selectivity, as demonstrated in the excitation efficiency plots shown in Figures 4(b), 4(d), and 4(f). A consistent pattern of high-fidelity control emerges across all targeted excitations. When the CF pulse is tuned to the complex pole corresponding to the first eigenmode (panel b), Qubit 1 is excited with high efficiency while the excitation of Qubit 2 and Qubit 3 remains negligible. Likewise, targeting the second pole (panel d) results in the clean and isolated excitation of only Qubit 2. The same high degree of selectivity is observed when targeting Qubit 3 (panel f), where it is exclusively excited with the other qubits remaining in their ground state. In all cases, the crosstalk to non-target qubits is suppressed to a level below ~10% of the target qubit's peak excitation at its maximum (at $t_0 = 0.2\ \mu s$) and generally stays within ~20% in the worst-case scenario. This represents a vast improvement over the Gaussian pulse method. For a quantitative comparison of Target Selectivity and Crosstalk Suppression Ratio for different pulse types in the lossless case, see *Supplementary Materials (S3)*.

The comparative analysis unequivocally demonstrates the superiority of CF pulses for high-fidelity qubit addressing in a realistic, nonlinear, multi-qubit circuit model. While the Gaussian pulse relies solely on frequency detuning for selectivity, its finite spectral bandwidth inevitably leads to off-resonant driving of neighboring modes in a coupled system. The CF pulse, by contrast, leverages both frequency and time-domain shaping. By matching the complex pole of a specific system eigenmode, the CF pulse excites a precise superposition of the system's basis states that channels energy into the target qubit while destructively interfering with the excitation pathways of other qubits. This results in both high efficiency (low reflection) and high selectivity (low crosstalk).

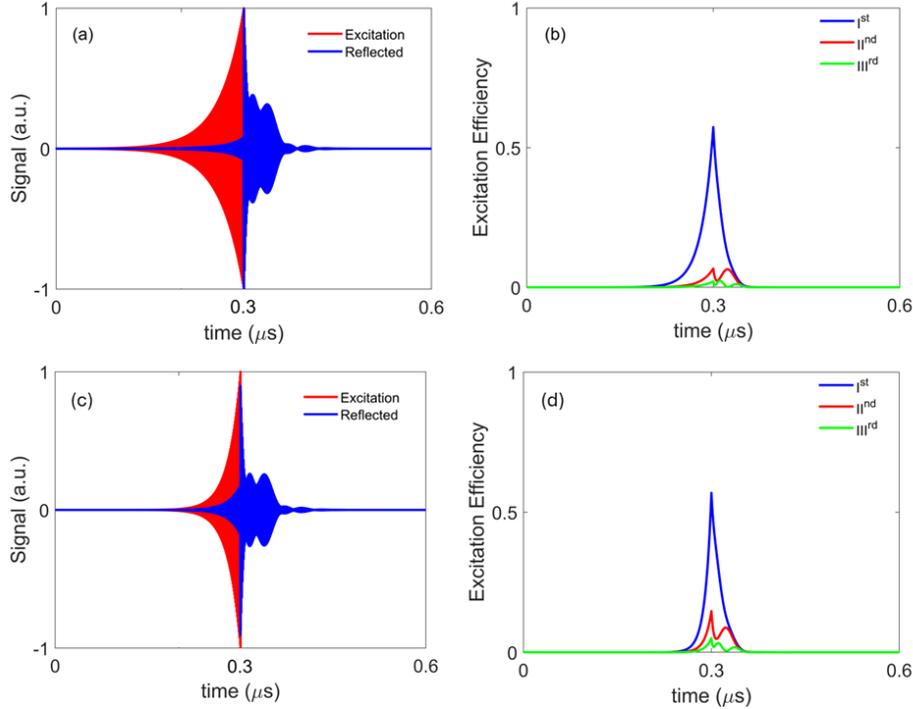

**Figure 5.** Selective qubit excitation in a three-qubit transmon system with intrinsic losses ($R_{shunt} = 0.2\,M\Omega$). (a) Time-domain profile of the input CF pulse (red) targeting Qubit 1's associated reflection zero and the significantly suppressed reflected signal (blue). (b) Corresponding excitation efficiencies ($\eta_1$ (blue), $\eta_2$ (red), $\eta_3$ (green)) demonstrating selective excitation of Qubit 1 with minimal crosstalk. (c) Time-domain profile of an input pulse (red) tuned to the complex conjugate of the reflection pole used in (a), showing substantial reflected signal (blue). (d) Corresponding excitation efficiencies, indicating reduced selectivity and increased crosstalk compared to excitation at the true reflection zero.

These ADS simulation results, which account for the inherent nonlinearity of the JJs, serve as a crucial validation of the principles derived from the theoretical model in Section II. They confirm that even in a realistic circuit, the underlying pole structure governs the system's response to tailored excitations, providing a powerful and practical method for achieving the precise quantum control necessary for scalable quantum information processing.

To assess the robustness of our CF pulse approach in a more realistic scenario, we incorporated intrinsic qubit losses into our model by introducing a shunt resistance ($R_{shunt} = 0.2\,M\Omega$) for each qubit. Figure 5 illustrates the system's response under these lossy conditions. When the system is driven with a CF pulse precisely tuned to a complex reflection zero associated predominantly with Qubit 1 (Fig. 5a,b), we observe a near-reflectionless regime, indicating efficient energy transfer into the system despite the presence of dissipation. While Qubit 1 is selectively excited to a significant level ($\eta_1 \approx 0.6$), approximately 30% of the total input energy is dissipated due to the intrinsic qubit losses, with negligible energy lost to reflection. In contrast, exciting the system at the complex conjugate of this reflection pole (Fig. 5c,d), a frequency one might naively consider based on time-reversal arguments in lossless systems, yields markedly different behavior. This excitation results in substantial reflection of the input pulse, signifying poor impedance matching. Although the peak excitation efficiency of Qubit 1 ($\eta_1 \approx 0.55$) is somewhat comparable to the optimal case, this comes at the cost of increased crosstalk to Qubits 2 and 3. Furthermore, the dominant energy loss mechanism in this latter scenario is reflection, accompanied by a comparatively lower, though still present, dissipative loss. These findings underscore the critical importance of targeting the true complex reflection zeros of the *lossy* system to achieve optimal absorption and maintain high selectivity, as further quantified by the selectivity and crosstalk suppression metrics detailed in *Supplementary Materials (S4)*.

## 4. Discussions

The successful demonstration of high-fidelity selective qubit excitation using CF pulses, as presented through both our theoretical framework and realistic ADS simulations of coupled superconducting transmon qubits, offers compelling prospects for advancing quantum processor capabilities. The core challenge in scaling quantum information systems lies not merely in increasing qubit numbers, but in maintaining and enhancing individual qubit control within an increasingly dense and interconnected environment. Our findings robustly suggest that CF pulse shaping, by targeting the specific complex zeros of the coupled system, provides a sophisticated and potent tool to navigate this intricate landscape.

A key implication for future quantum processor design is the enhanced spectral efficiency afforded by CF pulses. The demonstrated superior selectivity could significantly relax the often-stringent constraints on qubit frequency allocation. Current designs grapple with the "frequency budget," striving to place qubits far enough apart spectrally to minimize crosstalk while also avoiding regions prone to a high density of spurious modes or two-level system (TLS) defects. By effectively "seeing through" the complex coupling environment to address a target qubit via its unique pole, CF pulses might allow for more densely packed qubit arrays. This could lead to more compact chip layouts, reducing the overall device footprint and potentially mitigating challenges associated with long-range interconnects, differential thermal contraction, and maintaining mode uniformity across larger substrates. Such densification directly impacts the potential for more powerful quantum processors within a given physical size.

Furthermore, the impact on gate fidelities is profound. Control crosstalk is a dominant error source in multi-qubit systems, directly contributing to reduced single- and two-qubit gate performance. By substantially mitigating this crosstalk, as our results indicate, CF pulses can pave the way for intrinsically higher-fidelity operations. Achieving even incremental improvements in physical gate fidelities is crucial for pushing towards the operational thresholds required for effective quantum error correction (QEC). The ability to perform cleaner individual qubit rotations, without unintentionally perturbing neighbors, is foundational for constructing high-quality entangling gates and executing complex quantum algorithms.

The CF pulse technique may also streamline control hardware and calibration paradigms. While the generation of these pulses necessitates high-performance AWGs capable of precise amplitude and phase modulation, a requirement shared by other advanced control schemes like GRAPE or DRAG, the underlying physics of zero targeting offers a more deterministic design pathway. Given an accurately characterized system Hamiltonian (including qubit frequencies, anharmonicities, coupling strengths, and relevant loss rates), the optimal CF pulse parameters can, in principle, be calculated from the system's scattering matrix poles[42,60–63]. This contrasts favorably with purely numerical optimal control techniques, which can involve extensive, computationally intensive search procedures that may not always yield easily interpretable or robust pulse solutions. Reducing this calibration overhead is critical for managing the complexity of large-scale quantum systems. Moreover, the fundamental principles of matching excitation to system poles are not unique to transmon qubits or the specific capacitive coupling scheme investigated here. The CF pulse approach could potentially benefit other qubit modalities (e.g., flux qubits, spin qubits coupled to microwave resonators) and different coupling architectures where selective addressing in a strongly interacting environment is paramount.

Despite these promising implications, the experimental realization of CF pulse control presents several practical challenges that warrant careful consideration. Firstly, the fidelity of pulse generation is paramount. AWGs must possess sufficient bandwidth and resolution to accurately synthesize the often intricate temporal profiles of CF pulses, particularly the precise exponential decay components that define their imaginary frequency. Any distortions, phase noise, or bandwidth limitations in the control chain (from AWG to the qubit) could degrade the pulse shape, thereby diminishing the precision of pole targeting and reintroducing unwanted reflections or off-resonant driving. Secondly, the efficacy of the CF pulse technique is critically dependent on accurate system characterization. The locations of the complex zeros are sensitive functions of all system parameters: qubit frequencies, coupling strengths, resonator frequencies, and intrinsic and extrinsic loss rates. These parameters are not only subject to initial fabrication variations but can also drift due to environmental factors such as temperature fluctuations, stray fields, or even the long-term aging of materials. Robust and efficient protocols for initial system Hamiltonian tomography and subsequent tracking of parameter drifts will be essential for maintaining the high performance of CF pulses. The development of adaptive CF pulse schemes that can adjust to slow parameter variations in real-time would be a valuable future direction.

A third challenge relates to the inherent nonlinearity of Josephson junction-based qubits, such as transmons. Our ADS simulations demonstrate excellent performance in the quasi-linear regime, where drive powers are sufficiently low. However, the pursuit of faster gate operations often necessitates higher drive amplitudes. This can push the qubit into a more nonlinear regime where the effective potential landscape, and consequently the pole structure derived from a linearized model, may no longer accurately describe the system's response. The Kerr nonlinearity, for instance, makes the effective frequency of the qubit dependent on its own excitation state. Investigating the robustness of CF pulses to these drive-induced nonlinearities and determining the operational power limits for maintaining selectivity are crucial. While CF pulses are designed based on the linear response theory (poles of the scattering matrix), their performance in the presence of strong drives that can access the qubit's anharmonicity needs to be thoroughly explored.

Finally, the potential for leakage to non-computational states, particularly in weakly anharmonic transmons, must be considered. CF pulses, while tailored to specific poles, might possess broadband frequency components, especially if they are short. A careful analysis of the spectral content of CF pulses and their potential to drive transitions to higher excited states (e.g., the $|1\rangle \to |2\rangle$ transition) is necessary. This might require integrating CF pulse design with other leakage-reduction techniques, such as derivative-based methods (e.g., DRAG) applied to the CF pulse envelope or designing the CF pulse explicitly to minimize excitation at specific leakage transition frequencies.

## 5. Conclusions

This work systematically investigated the application of complex frequency (CF) pulses for high-fidelity, selective control of individual qubits in coupled, multi-qubit superconducting systems, a critical challenge for scalable quantum computation. By tailoring excitation pulses to the specific complex reflection zeros of the coupled qubit-waveguide system, our approach directly mitigates control crosstalk. We developed a theoretical framework using an input-output approach and linearized Heisenberg equations, which guided the design of CF pulses by identifying the system's characteristic complex poles. This concept was then rigorously tested through comprehensive transient simulations in Keysight's Advanced Design System (ADS), employing a realistic nonlinear model for three Josephson junction-based transmon qubits. The central result of our investigation is the demonstrated superior performance of CF pulses over conventional Gaussian pulses. ADS simulations revealed that CF pulses achieve highly selective excitation of a target qubit, with crosstalk to neighboring qubits suppressed to ~10% of the target qubit's peak excitation, even in spectrally crowded scenarios. In contrast, Gaussian pulses exhibited substantial crosstalk, reaching up to 35% This enhanced performance stems from the CF pulse's ability to effectively impedance match to the target eigenmode, leading to near-perfect energy absorption and minimal reflection. The significant crosstalk mitigation achieved with CF pulses offers a promising route to higher-fidelity quantum operations, potentially enabling denser qubit architectures and simplified control calibration in future quantum processors. While experimental implementation will require precise pulse synthesis and system characterization, our theoretical and simulation results strongly advocate for CF pulse shaping as a potent tool for advancing quantum control. Future efforts will be directed towards experimental validation and assessing the robustness of this technique against various noise sources and system imperfections.

**Notes**

The authors declare no competing financial interest.

**Acknowledgments**

The authors acknowledge financial support from the DoE and AFOSR.

**Supporting Information**

A detailed derivation of the system's equations of motion and demonstration that conventional Gaussian pulses result in significant crosstalk. Quantitative metrics, Target Selectivity and Crosstalk Suppression Ratio, reveal Complex Frequency (CF) pulses for lossless and lossy systems.

25. Chiaro, B. & Zhang, Y. Active Leakage Cancellation in Single Qubit Gates. Preprint at https://doi.org/10.48550/arXiv.2503.14731 (2025).

26. Hyyppä, E. *et al.* Reducing Leakage of Single-Qubit Gates for Superconducting Quantum Processors Using Analytical Control Pulse Envelopes. *PRX Quantum* **5**, 030353 (2024).

27. Zong, Z. *et al.* Optimization of a Controlled- Z Gate with Data-Driven Gradient-Ascent Pulse Engineering in a Superconducting-Qubit System. *Phys. Rev. Appl.* **15**, 064005 (2021).

28. Matsuda, R. *et al.* Selective Excitation of Superconducting Qubits with a Shared Control Line through Pulse Shaping. Preprint at https://doi.org/10.48550/arXiv.2501.10710 (2025).

29. Werninghaus, M. *et al.* Leakage reduction in fast superconducting qubit gates via optimal control. *Npj Quantum Inf.* **7**, 14 (2021).

30. Motzoi, F. & Wilhelm, F. K. Improving frequency selection of driven pulses using derivative-based transition suppression. *Phys. Rev. A* **88**, 062318 (2013).

31. Chen, Z.-J. *et al.* Robust and optimal control of open quantum systems. *Sci. Adv.* **11**, eadr0875 (2025).

32. Ansel, Q., Fischer, J., Sugny, D. & Bellomo, B. Optimal control and selectivity of qubits in contact with a structured environment. *Phys. Rev. A* **106**, 043702 (2022).

33. Magann, A. B. *et al.* From Pulses to Circuits and Back Again: A Quantum Optimal Control Perspective on Variational Quantum Algorithms. *PRX Quantum* **2**, 010101 (2021).
27

# Supplementary Materials
# Selective Addressing of Coupled Qubits via Complex Frequency Zero Targeting

Deepanshu Trivedi[1], Laraib Niaz[1], and Alex Krasnok[1,2]

[1]Department of Electrical Engineering, Florida International University, 33174, Miami, USA

[2]Knight Foundation School of Computing and Information Sciences, Florida International University, 33174, Miami, USA

To whom correspondence should be addressed: akrasnok@fiu.edu

## S1. Derivation of System Equations of Motion

This section provides a more detailed derivation of the coupled Heisenberg equations of motion presented in Eqs. (1-3) of the main text. The model describes a system of three two-level qubits interacting with a single resonator mode, which itself is coupled to an input-output waveguide and potentially to an internal loss channel. Each qubit is also coupled to its own independent environment leading to spontaneous emission.

The total Hamiltonian $\hat{H}_{\text{total}}$ for the system and its environment can be written as: $\hat{H}_{\text{total}} = \hat{H}_{\text{sys}} + \hat{H}_{\text{bath}} + \hat{H}_{\text{sys-bath}}$. The system Hamiltonian $\hat{H}_{\text{sys}}$ consists of the resonator, the qubits, and their mutual interaction: $\hat{H}_{\text{sys}} = \hat{H}_{\text{res}} + \sum_{j=1}^{3} \hat{H}_{\text{qubit},j} + \sum_{j=1}^{3} \hat{H}_{\text{int},j}$ where:

- The resonator Hamiltonian is $\hat{H}_{\text{res}} = \hbar \omega_c \hat{a}^\dagger \hat{a}$, with $\hat{a}$ ($\hat{a}^\dagger$) being the annihilation (creation) operator for the resonator mode of frequency $\omega_c$.

- The Hamiltonian for the $j$-th qubit is $\hat{H}_{\text{qubit},j} = \frac{1}{2}\hbar \omega_{aj} \hat{\sigma}_z^{(j)}$, where $\omega_{aj}$ is the transition frequency of qubit $j$, and $\hat{\sigma}_z^{(j)}$ is the Pauli-Z operator. We define $\hat{\sigma}_-^{(j)}$ and $\hat{\sigma}_+^{(j)}$ as the corresponding lowering and raising operators.

- The interaction Hamiltonian between the *j*-th qubit and the resonator mode is given by the Tavis-Cummings model under the rotating wave approximation (RWA): $\hat{H}_{\text{int},j} = \hbar g_j (\hat{a}^\dagger \hat{\sigma}_-^{(j)} + \hat{a} \hat{\sigma}_+^{(j)})$, where $g_j$ is the coupling strength.

The bath Hamiltonian $\hat{H}_{\text{bath}}$ and system-bath interaction $\hat{H}_{\text{sys-bath}}$ include contributions from the waveguide coupled to the resonator, internal loss mechanisms for the resonator, and local environments for each qubit causing spontaneous emission: $\hat{H}_{\text{bath}} = \hat{H}_{\text{bath,wg}} + \hat{H}_{\text{bath,res-int}} + \sum_{j=1}^{3} \hat{H}_{\text{bath,q},j}$, $\hat{H}_{\text{sys-bath}} = \hat{H}_{\text{sys-wg}} + \hat{H}_{\text{sys-res-int}} + \sum_{j=1}^{3} \hat{H}_{\text{sys-q,env},j}$ Specifically:

- For the resonator-waveguide coupling: $\hat{H}_{\text{bath,wg}} = \sum_k \hbar \omega_k \hat{b}_k^\dagger \hat{b}_k$ and $\hat{H}_{\text{sys-wg}} = i\hbar \sum_k \lambda_k (\hat{a}^\dagger \hat{b}_k - \hat{a} \hat{b}_k^\dagger)$.

- For internal resonator losses (if $1/\tau_i \neq 0$): $\hat{H}_{\text{bath,res-int}} = \sum_m \hbar \omega_m \hat{c}_m^\dagger \hat{c}_m$ and $\hat{H}_{\text{sys-res-int}} = i\hbar \sum_m \mu_m (\hat{a}^\dagger \hat{c}_m - \hat{a} \hat{c}_m^\dagger)$.

- For qubit spontaneous emission: $\hat{H}_{\text{bath,q},j} = \sum_l \hbar \omega_{jl} \hat{d}_{jl}^\dagger \hat{d}_{jl}$ and $\hat{H}_{\text{sys-q,env},j} = i\hbar \sum_l \nu_{jl} (\hat{\sigma}_+^{(j)} \hat{d}_{jl} - \hat{\sigma}_-^{(j)} \hat{d}_{jl}^\dagger)$.

The time evolution of any system operator $\hat{O}$ is given by the Heisenberg equation: $\frac{d\hat{O}}{dt} = \frac{1}{i\hbar}[\hat{O}, \hat{H}_{\text{total}}]$ (assuming $\hat{O}$ has no explicit time dependence). We apply this to derive the equations for $\hat{a}$, $\hat{\sigma}_-^{(j)}$, and $\hat{\sigma}_z^{(j)}$.

Using the commutation relation $[\hat{a}, \hat{a}^\dagger \hat{a}] = \hat{a}$ and $[\hat{a}, \hat{a}\hat{\sigma}_+^{(j)}] = 0$, $[\hat{a}, \hat{a}^\dagger \hat{\sigma}_-^{(j)}] = \hat{\sigma}_-^{(j)}$, we get: $\frac{1}{i\hbar}[\hat{a}, \hat{H}_{\text{sys}}] = i\omega_c \hat{a} - i\sum_j g_j \hat{\sigma}_-^{(j)}$ The interaction with the waveguide bath $\hat{H}_{\text{sys-wg}}$ is treated using standard input-output theory(*1*). Assuming a Markovian bath, this coupling introduces a decay term and an input field term. The decay rate into the waveguide is $\kappa_r/2$. If we define the amplitude decay time $\tau_r$ such that $1/\tau_r = \kappa_r/2$, the decay term becomes $(1/\tau_r)\hat{a}$. The input field term is $\sqrt{\kappa_r} \hat{b}_{in}(t) = \sqrt{2/\tau_r} \hat{b}_{in}(t)$, where $\hat{b}_{in}(t)$ is the input field operator from the waveguide. This matches the $k_r \hat{b}_{in}(t)$ term in Eq. (1) of the main text if $\kappa_r = \sqrt{2/\tau_r}$ and $b_{in}(t)$ is the expectation value of the (normalized) input field operator. Similarly, the coupling to an internal loss channel for the resonator, $\hat{H}_{\text{sys-res-int}}$, yields a decay term $(1/\tau_i)\hat{a}$, where $1/\tau_i = \kappa_i/2$ is the amplitude decay rate due to internal losses.

Combining these, the Heisenberg-Langevin equation for $\hat{a}$ becomes: $\frac{d\hat{a}}{dt} = i\omega_c \hat{a} - i\sum_j g_j \hat{\sigma}_-^{(j)} - \left(\frac{1}{\tau_r} + \frac{1}{\tau_i}\right)\hat{a} + k_r \hat{b}_{in}(t) + \hat{F}_r(t) + \hat{F}_i(t)$ where $\hat{F}_r(t)$ and $\hat{F}_i(t)$ are noise operators associated with the external and internal baths, respectively. Taking the expectation value $\langle \cdot \rangle$ and assuming $\langle \hat{F}_r(t) \rangle = \langle \hat{F}_i(t) \rangle = 0$ (which is standard for vacuum or coherent input states when considering the evolution of expectation values), we arrive at Eq. (1) of the main text:

$$\frac{d}{dt}\langle a \rangle = \left(i\omega_c - \frac{1}{\tau_r} - \frac{1}{\tau_i}\right)\langle a \rangle - i\sum_j g_j \langle \sigma_-^{(j)} \rangle + k_r b_{in}(t)$$

Next, using the Pauli operator commutation relations $[\hat{\sigma}_-^{(j)}, \hat{\sigma}_z^{(j)}] = 2\hat{\sigma}_-^{(j)}$ and $[\hat{\sigma}_-^{(j)}, \hat{\sigma}_+^{(j)}] = -\hat{\sigma}_z^{(j)}$ (standard definitions where $\hat{\sigma}_z = \begin{pmatrix} 1 & 0 \\ 0 & -1 \end{pmatrix}$), we find:

- $\frac{1}{i\hbar}[\hat{\sigma}_-^{(j)}, \hat{H}_{\text{qubit},j}] = \frac{1}{i\hbar}[\hat{\sigma}_-^{(j)}, \frac{1}{2}\hbar\omega_{aj}\hat{\sigma}_z^{(j)}] = i\omega_{aj}\hat{\sigma}_-^{(j)}$.

- $\frac{1}{i\hbar}[\hat{\sigma}_-^{(j)}, \hat{H}_{\text{int},j}] = \frac{1}{i\hbar}[\hat{\sigma}_-^{(j)}, \hbar g_j(\hat{a}^\dagger \hat{\sigma}_-^{(j)} + \hat{a}\hat{\sigma}_+^{(j)})] = ig_j \hat{a}[\hat{\sigma}_-^{(j)}, \hat{\sigma}_+^{(j)}] = ig_j \hat{a}\hat{\sigma}_z^{(j)}$.

The coupling of qubit $j$ to its local environment leads to spontaneous emission. Under the Born-Markov approximation, this contributes a decay term $(\Gamma_s^{(j)}/2)\hat{\sigma}_-^{(j)}$ to the equation for $\hat{\sigma}_-^{(j)}$, where $\Gamma_s^{(j)}$ is the energy decay rate (or $1/T_1$) for qubit $j$. This term describes the loss of coherence. Thus, the Heisenberg-Langevin equation for $\hat{\sigma}_-^{(j)}$ is: $\frac{d\hat{\sigma}_-^{(j)}}{dt} = i\omega_{aj}\hat{\sigma}_-^{(j)} + ig_j \hat{a}\hat{\sigma}_z^{(j)} - \frac{\Gamma_s^{(j)}}{2}\hat{\sigma}_-^{(j)} + \hat{F}_q^{(j)}(t)$ where $\hat{F}_q^{(j)}(t)$ is the corresponding noise operator. Taking the expectation value, assuming $\langle \hat{F}_q^{(j)}(t) \rangle = 0$, and taking expectation values of products like $\langle \hat{a}\hat{\sigma}_z^{(j)} \rangle \approx \langle a \rangle \langle \sigma_z^{(j)} \rangle$, we arrive at:

$$\frac{d}{dt}\langle \sigma_-^{(j)} \rangle = \left(i\omega_{aj} - \frac{\Gamma_s^{(j)}}{2}\right)\langle \sigma_-^{(j)} \rangle + ig_j \langle a \rangle \langle \sigma_z^{(j)} \rangle$$

Finally, using $[\hat{\sigma}_z^{(j)}, \hat{\sigma}_-^{(j)}] = -2\hat{\sigma}_-^{(j)}$ and $[\hat{\sigma}_z^{(j)}, \hat{\sigma}_+^{(j)}] = 2\hat{\sigma}_+^{(j)}$:

- $\frac{1}{i\hbar}[\hat{\sigma}_z^{(j)}, \hat{H}_{\text{qubit},j}] = 0.$

- $\frac{1}{i\hbar}[\hat{\sigma}_z^{(j)}, \hat{H}_{\text{int},j}] = \frac{1}{i\hbar}[\hat{\sigma}_z^{(j)}, \hbar g_j(\hat{a}^\dagger \hat{\sigma}_-^{(j)} + \hat{a}\hat{\sigma}_+^{(j)})] = ig_j(\hat{a}^\dagger[\hat{\sigma}_z^{(j)}, \hat{\sigma}_-^{(j)}] + \hat{a}[\hat{\sigma}_z^{(j)}, \hat{\sigma}_+^{(j)}]) =$ $ig_j(\hat{a}^\dagger(-2\hat{\sigma}_-^{(j)}) + \hat{a}(2\hat{\sigma}_+^{(j)})) = 2ig_j(-\hat{a}^\dagger \hat{\sigma}_-^{(j)} + \hat{a}\hat{\sigma}_+^{(j)}).$

The spontaneous emission process causes the qubit to decay from the excited state to the ground state. This relaxation is described by the term $-\Gamma_s^{(j)}(\hat{\sigma}_z^{(j)} + \hat{I})$, where $\hat{I}$ is the identity operator. This term ensures that in the absence of driving, $\langle \hat{\sigma}_z^{(j)} \rangle$ decays towards $-1$ (the ground state) with a rate $\Gamma_s^{(j)}$. This form is standard from the Lindblad master equation for a spontaneously emitting two-level system.

Combining these, the Heisenberg-Langevin equation for $\hat{\sigma}_z^{(j)}$ is: $\frac{d\hat{\sigma}_z^{(j)}}{dt} = 2ig_j(-\hat{a}^\dagger \hat{\sigma}_-^{(j)} + \hat{a}\hat{\sigma}_+^{(j)}) - \Gamma_s^{(j)}(\hat{\sigma}_z^{(j)} + \hat{I}) + \hat{G}_q^{(j)}(t)$ where $\hat{G}_q^{(j)}(t)$ is the noise operator. Taking the expectation value, assuming $\langle \hat{G}_q^{(j)}(t) \rangle = 0$, and often factorizing terms such as $\langle \hat{a}^\dagger \hat{\sigma}_-^{(j)} \rangle \approx \langle a^\dagger \rangle \langle \sigma_-^{(j)} \rangle$, we obtain:

$$\frac{d}{dt}\langle \sigma_z^{(j)} \rangle = 2ig_j(-\langle a^\dagger \rangle \langle \sigma_-^{(j)} \rangle + \langle a \rangle \langle \sigma_+^{(j)} \rangle) - \Gamma_s^{(j)}(\langle \sigma_z^{(j)} \rangle + 1)$$

This derivation relies on standard approximations in quantum optics, namely the Born-Markov approximation for treating system-bath interactions and the rotating wave approximation for the qubit-resonator coupling. The resulting equations for the expectation values form a closed set, often referred to as generalized Bloch equations when coupled with the field equation, providing a powerful semi-classical framework for analyzing system dynamics.

## S2. Gaussian Pulse Excitation in the Theoretical Model

To provide a direct comparison with the Complex Frequency (CF) pulse performance presented in the main text (Figure 2), this section details the system's response to conventional Gaussian pulse excitation within the same theoretical framework. The dynamics are derived from the Heisenberg equations of motion (Eqs. 1-3 in the main text) for the three-qubit, one-resonator system. For these simulations, Gaussian-enveloped microwave pulses are tuned to the real resonance frequencies corresponding to the eigenmodes primarily associated with each target qubit. The energy content of these Gaussian pulses is set to be comparable to that of the CF pulses used in the main text to ensure a fair comparison of their efficacy.

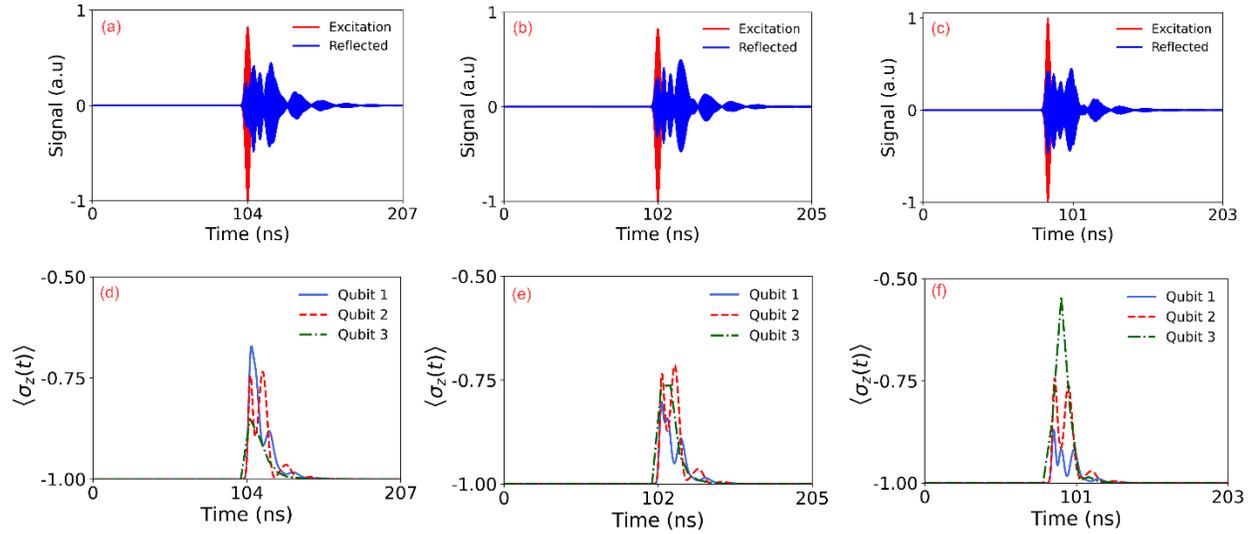

**Figure S5.** Theoretical simulation of qubit population dynamics under Gaussian pulse excitation. Gaussian pulses, with energy comparable to the CF pulses (Fig. 2), are tuned to the real resonance frequency of the eigenmode predominantly associated with: (a) Qubit 1, (b) Qubit 2, and (c) Qubit 3. (d-f) Population inversion $\langle \sigma_z^{(j)}(t) \rangle$ for Qubit 1 (blue solid line), Qubit 2 (red dashed line), and Qubit 3 (green dotted line) as a function of time, calculated using the theoretical model. Significant crosstalk to non-target qubits is evident in all cases.

Figure S1 illustrates the simulated population inversion, $\langle \sigma_z^{(j)}(t) \rangle$, for all three qubits when attempting to selectively excite each one using a Gaussian pulse. When targeting Qubit 1 (panel a), Qubit 1 (solid blue line) shows the intended excitation. However, significant unwanted excitation, or crosstalk, is observed in Qubit 2 (dashed red line) and, to a lesser extent, in Qubit 3 (dotted green line). Similarly, when the Gaussian pulse is tuned to excite Qubit 2 (panel b), Qubit 2 is excited, but substantial crosstalk is induced in both Qubit 1 and Qubit 3. Targeting Qubit 3 (panel c) also results in its excitation but is accompanied by considerable off-resonant driving of Qubits 1 and 2. These results clearly indicate that, within this theoretical model, standard Gaussian pulses lead to significant spectral overlap and consequential crosstalk in a system with closely spaced and coupled qubits. The lack of selectivity underscores the limitations of conventional pulse techniques and highlights the necessity for more advanced methods, such as the CF pulse approach, to achieve high-fidelity individual qubit control.

## S3. Target Selectivity and Crosstalk Suppression

This supplementary section provides a detailed quantitative analysis of the selective excitation performance using the figures of merit: Target Selectivity ($S_i^{target}$) and Crosstalk Suppression Ratio ($C_i$). These metrics are evaluated at the end of the excitation pulse.

**Target Selectivity ($S_i^{target}$):** This metric is defined as the fraction of energy stored in the intended target qubit $i$ relative to the total energy stored across all qubits in the system: $S_i^{target} = \frac{\eta_i}{\sum_{k=1}^{N} \eta_k}$ where $\eta_k$ is the excitation efficiency (proportional to energy stored) of the $k$-th qubit, and $N$ is the total number of qubits. A value of $S_i^{target}$ close to 1 indicates that the excitation pulse successfully directed most of the stored energy into the desired target qubit.

**Crosstalk Suppression Ratio ($C_i$):** This metric quantifies the isolation of the target qubit and is defined as the ratio of the energy stored in the target qubit $i$ to the maximum energy stored in any single non-target qubit: $C_i = \frac{\eta_i}{\max_{k \neq i}(\eta_k)}$ A higher value of $C_i$ signifies better suppression of unwanted energy transfer to non-target qubits, indicating superior addressability.

In the lossless scenario, we compare the performance of conventional Gaussian pulses, Gaussian pulses tuned to the real part of the system's complex reflection zeros, and the proposed CF pulses.

**Conventional Gaussian Pulse Excitation:** This table presents the individual qubit efficiencies ($\eta_1, \eta_2, \eta_3$) when targeting Qubit 1, Qubit 2, or Qubit 3 using standard Gaussian pulses whose carrier frequencies are tuned to the respective qubit's nominal resonance frequency. For instance, when targeting Qubit 1, $\eta_1 = 50.8\%$, but significant crosstalk occurs, with $\eta_2 = 32.7\%$ and $\eta_3 = 2.4\%$. Similar substantial crosstalk is observed when targeting Qubit 2 ($\eta_1 = 15.4\%, \eta_2 = 45.2\%, \eta_3 = 20.3\%$) and Qubit 3 ($\eta_1 = 4.1\%, \eta_2 = 20.0\%, \eta_3 = 48.9\%$). This highlights the limitations of conventional Gaussian pulses in spectrally crowded systems.

| Excitation (**Gaussian**) | Efficiency (%) | | |
|---|---|---|---|
| | $\eta_1$ | $\eta_2$ | $\eta_3$ |
| Qubit 1 | 50.8 | 32.7 | 2.4 |
| Qubit 2 | 15.4 | 45.2 | 20.3 |
| Qubit 3 | 4.1 | 20 | 48.9 |

**Gaussian Pulses Tuned to System Zeros vs. CF Pulses:** The subsequent tables compare Gaussian pulses (with carrier frequencies at $\text{Re}(\omega_z)$) and CF pulses when targeting the system's complex reflection zeros (denoted Leftmost, Middle, and Rightmost, corresponding primarily to addressing Qubit 3, Qubit 2, and Qubit 1, respectively).

**Target Selectivity ($\mathcal{S}_i^{target}$):** When using Gaussian pulses tuned to the real part of the zeros, selectivity remains modest. For example, targeting the "Rightmost zero" (associated with Qubit 1) yields $\mathcal{S}_1^{target} = 0.59$. In stark contrast, CF pulses demonstrate significantly enhanced selectivity. Targeting the "Rightmost zero" with a CF pulse results in $\mathcal{S}_1^{target} = 0.91$. Similarly, for the "Middle zero" (Qubit 2 target), CF pulses achieve $\mathcal{S}_2^{target} = 0.81$ (compared to 0.56 for Gaussian), and for the "Leftmost zero" (Qubit 3 target), $\mathcal{S}_3^{target} = 0.85$ (compared to 0.67 for Gaussian).

| Excitation at frequency of zero (**Gaussian** pulse) | Target Selectivity ($\mathcal{S}_i^{target}$) | | |
|---|---|---|---|
| | Qubit 1 | Qubit 2 | Qubit 3 |
| Leftmost zero | 0.06 | 0.27 | 0.67 |
| Middle zero | 0.19 | 0.56 | 0.25 |
| Rightmost zero | 0.59 | 0.38 | 0.03 |

| Excitation at frequency of zero | Target Selectivity ($\mathcal{S}_i^{target}$) |
|---|---|

| (**Complex Frequency** pulse) | | | |
|---|---|---|---|
|  | Qubit 1 | Qubit 2 | Qubit 3 |
| Leftmost zero | 0.02 | 0.12 | 0.85 |
| Middle zero | 0.09 | 0.81 | 0.09 |
| Rightmost zero | 0.91 | 0.07 | 0.02 |

**Crosstalk Suppression Ratio ($\mathcal{C}_i$):** Gaussian pulses tuned to the real part of the zeros show limited crosstalk suppression. For the "Rightmost zero" (Qubit 1 target), $\mathcal{C}_1 = 1.55$. CF pulses achieve vastly superior crosstalk suppression. For the "Rightmost zero," $\mathcal{C}_1 = 13.07$. Targeting the "Middle zero" (Qubit 2) yields $\mathcal{C}_2 = 8.37$ (compared to 2.23 for Gaussian), and for the "Leftmost zero" (Qubit 3), $\mathcal{C}_3 = 7.08$ (compared to 2.445 for Gaussian).

| Excitation at frequency of zero (**Gaussian pulse**) | Crosstalk Suppression Ratio ($\mathcal{C}_i$) | | |
|---|---|---|---|
|  | Qubit 1 | Qubit 2 | Qubit 3 |
| Leftmost zero | 0.08 | 0.41 | 2.445 |
| Middle zero | 0.34 | 2.23 | 0.45 |
| Rightmost zero | 1.55 | 0.64 | 0.05 |

| Excitation at frequency of zero (**Complex Frequency**) | Crosstalk Suppression Ratio ($\mathcal{C}_i$) | | |
|---|---|---|---|
|  | Qubit 1 | Qubit 2 | Qubit 3 |
| Leftmost zero | 0.03 | 0.15 | 7.08 |
| Middle zero | 0.12 | 8.37 | 0.10 |
| Rightmost zero | 13.07 | 0.08 | 0.02 |

The final efficiency table for CF pulse excitation directly shows the outcome: when targeting Qubit 1 (Rightmost zero), $\eta_1 = 85.0\%$, with $\eta_2 = 6.5\%$ and $\eta_3 = 1.7\%$. When targeting Qubit 2 (Middle zero), $\eta_2 = 77.0\%$, with $\eta_1 = 9.2\%$ and $\eta_3 = 8.4\%$. When targeting Qubit 3 (Leftmost zero), $\eta_3 = 80.7\%$, with $\eta_1 = 2.3\%$ and $\eta_2 = 11.4\%$. These values clearly indicate the high efficiency and low crosstalk achieved by CF pulses in the lossless case.

| Excitation (**Complex Frequency**) | Efficiency (%) | | |
|---|---|---|---|
| | $\eta_1$ | $\eta_2$ | $\eta_3$ |
| Qubit 1 | 85 | 6.5 | 1.7. |
| Qubit 2 | 9.2 | 77 | 8.4 |
| Qubit 3 | 2.3 | 11.4 | 80.7 |

## S4. Target Selectivity and Crosstalk Suppression in Lossy case

This section presents data for the system including intrinsic qubit losses ($R_{shunt} = 0.2\,M\Omega$). The tables focus on targeting Qubit 1 (via the "Rightmost Zero" of the lossy system) and compare the performance of a CF pulse tuned to this true reflection zero against a pulse tuned to its complex conjugate.

**Individual Qubit Efficiencies ($\eta_i$):**

o Exciting at the true reflection zero (Rightmost Zero): $\eta_1 = 56.6\%$, $\eta_2 = 6.7\%$, $\eta_3 = 2.1\%$. The sum of these efficiencies ($\sim 65.4\%$) is less than 100%, with the remainder primarily lost to dissipation, as discussed in the main text.

o Exciting at the complex conjugate of the pole: $\eta_1 = 55.8\%$, $\eta_2 = 14.5\%$, $\eta_3 = 5.0\%$. The sum ($\sim 75.3\%$) is higher, but as shown in the main text (Fig. 6c), this case involves significant reflection losses.

| *Excitation at frequency of zero* | Efficiency (%) | | |
|---|---|---|---|
| | $\eta_1$ | $\eta_2$ | $\eta_3$ |
| Rightmost Zero | 56.6 | 6.7 | 2.1 |

| *Excitation at frequency of complex conjugate of pole* | Efficiency (%) | | |
|---|---|---|---|
| | $\eta_1$ | $\eta_2$ | $\eta_3$ |
| Rightmost Zero | 55.8 | 14.5 | 5. |

**Target Selectivity ($S_i^{target}$):**

o Exciting at the true reflection zero (targeting Q1): $S_1^{target} = 0.87$, with $S_2^{target} = 0.10$ and $S_3^{target} = 0.03$.

o Exciting at the complex conjugate of the pole (targeting Q1): $S_1^{target} = 0.74$, with $S_2^{target} = 0.19$ and $S_3^{target} = 0.07$. This shows that targeting the true zero yields higher selectivity, meaning a larger fraction of the energy *stored within the qubits* resides in the target qubit.

| Excitation at frequency of zero | Target Selectivity ($S_i^{target}$) | | |
|---|---|---|---|
| | Qubit 1 | Qubit 2 | Qubit 3 |
| Rightmost Zero | 0.87 | 0.10 | 0.03 |

| Excitation at frequency of complex conjugate of pole | Target Selectivity ($S_i^{target}$) | | |
|---|---|---|---|
| | Qubit 1 | Qubit 2 | Qubit 3 |
| Rightmost Zero | 0.74 | 0.19 | 0.07 |

**Crosstalk Suppression ($C_i$ and related ratios):** For these tables, where the "Rightmost Zero" is consistently targeted (i.e., Qubit 1 is the target), the "Qubit 1" column shows the Crosstalk Suppression Ratio $C_1 = \eta_1/\max(\eta_2, \eta_3)$. The "Qubit 2" and "Qubit 3" columns show the ratios $\eta_2/\eta_1$ and $\eta_3/\eta_1$ respectively, quantifying the excitation of non-target qubits relative to the target.

- Exciting at the true reflection zero: $C_1 = 8.45$. The relative excitations are $\eta_2/\eta_1 \approx 0.118$ and $\eta_3/\eta_1 \approx 0.037$.
- Exciting at the complex conjugate of the pole: $C_1 = 3.84$. The relative excitations are $\eta_2/\eta_1 \approx 0.260$ and $\eta_3/\eta_1 \approx 0.090$.

| Excitation at frequency of zero | Crosstalk Suppression Ratio () | | |
|---|---|---|---|
| | Qubit 1 | Qubit 2 | Qubit 3 |
| Rightmost Zero | 8.45 | 0.12 | 0.04 |

| Excitation at frequency of complex conjugate of pole | Crosstalk Suppression Ratio () | | |
|---|---|---|---|
| | Qubit 1 | Qubit 2 | Qubit 3 |
| Rightmost Zero | 3.84 | 0.26 | 0.09 |

**These values clearly demonstrate that exciting at the true reflection zero of the lossy system results in significantly better crosstalk suppression ($C_1$ is more than doubled) and lower relative excitation of non-target qubits compared to exciting at the complex conjugate frequency.** This quantitatively supports the findings presented in Figure 6 of the main text, emphasizing the importance of precise CF pulse design for lossy systems.